\documentclass{JHEP3}
\usepackage{amsmath,amssymb,amsfonts}
\usepackage{epsfig,nicefrac}
\newcommand{\oao}[2]{{#1\atopwithdelims[]#2}}
\def\zi{\mathbb{Z}}
\def\di{\text{d}}
\def\rrangle{\rangle \hskip-.5mm \rangle}
\def\hl{\frac{1}{2}}
\def\cn{\mathcal{N}}

\title{ Comments on geometric and universal open string tachyons near fivebranes
}

\author{
        Dan Isra\"el\\
   \\ \textsc{greco}, Institut d'Astrophysique de Paris\footnote{Unit\'e mixte de Recherche
7095, CNRS -- Universit\'e Pierre et Marie Curie},
98bis Bd Arago,75014 Paris France \\
        E-mail:  \email{israel@iap.fr}
}

\abstract{
In a recent work~\cite{Sen:2007cz}, Sen studied unstable D-branes in NS5-brane backgrounds
and argued that in the strong curvature regime  the
universal open string tachyon (on D-branes of the wrong dimensionality) and the geometric
tachyon (on D-branes that are \textsc{bps} in flat space but not in this background) may become equivalent.
We study in this note  an example of a non-\textsc{bps}
suspended D-brane vs. a \textsc{bps} D-brane at equal distance between two
fivebranes. We use  boundary worldsheet \textsc{cft} methods to show that these
two unstable branes are identical.
}

\preprint{hep-th/0703261}


\begin{document}

\section{Introduction}
Non-\textsc{bps} brane configurations, e.g. superstring D-branes with the wrong dimensionality,
have attracted lot of interest in recent years due to their
relevance in supersymmetry breaking and cosmology (see~\cite{Sen:2004nf} and references therein).

In~\cite{Kutasov:2004ct} Kutasov studied \textsc{bps} D-branes near an NS5-brane stack with a transverse circle. If
the D-brane and the NS5-branes are at antipodal points, the D-brane sits at a maximum of its effective
potential as it will tend to fall onto the fivebranes. Its open string spectrum has a
{\it geometric tachyon}, even though far from the NS5-branes the D-brane is \textsc{bps}. Interestingly, the
properties of this geometrical tachyon are closely related to those of the {\it universal tachyon} on
non-\textsc{bps} branes.

Recently Sen considered further configurations of \textsc{bps} and non-\textsc{bps} D-branes
near NS5-branes~\cite{Sen:2007cz}. He proposed that, in the limit where the number of fivebranes
is small (i.e. strong curvature
in target space) one can identify the geometric tachyon on some \textsc{bps} branes with the
universal tachyon on other non-\textsc{bps} branes. One example involved on the one hand a \textsc{bps}
brane halfway between two NS5-branes (called G-type) and on the other hand a non-\textsc{bps} brane
stretched between two fivebranes (called S-type). However in this regime
all $\alpha '$ corrections (perturbative and non-perturbative) play a role.

In this note we provide an exact worldsheet \textsc{cft} description of this phenomena.
We start with a slightly different  system, a ring of $k$ NS5-branes with a non-compact transverse space.
D-branes in the near-horizon geometry of the fivebranes ring are known~\cite{Eguchi:2004ik,Israel:2005fn}.
Among them, one has  S-type branes stretched between any pair of fivebranes, and G-type branes whose worldvolume
fill a disk at the center of the ring.

In the particular case $k=2$ it reduces to a system closely related to the model studied by Sen --~the
near-horizon geometry of a pair of NS5-branes. The brane moduli space becomes smaller, leaving one S-type brane
stretched between the NS5-branes and one G-type brane which is a fuzzy point halfway between the fivebranes in the transverse space.
For these particular branes, the transverse circle is irrelevant. Therefore we expect to
find the same phenomena as in Sen's analysis.

Borrowing the technical details from~\cite{Israel:2005fn} we find that for $k=2$ the G-type
brane and the S-type brane are identified. They have the same one-point function on the disc and their 
open string spectrum is identical, built on  a tachyon with $M^2=-\nicefrac{1}{2\alpha'}$.

In sect.~\ref{ring} we analyze the two sorts of D-branes in the ring geometry. Then in sect.~\ref{pair} we focus
on the pair of fivebranes, and show that the two unstable branes coincide. We compare finally our results
with those of~\cite{Sen:2007cz}. Some useful material is gathered in two appendices.

\section{D-branes near a ring of fivebranes}
\label{ring}
A supersymmetric  superstring background is obtained by spreading
$k$ parallel NS5-branes on a circle in their non-compact transverse space. Calling $R$ its radius
in string units, one can define a double scaling limit of the system~\cite{Giveon:1999px}
($g_s \to 0$, $R/g_s$ fixed) in order to reach the near-horizon geometry of all  fivebranes
simultaneously. Remarkably the ring geometry is an exactly
solvable worldsheet \textsc{cft}~\cite{Sfetsos:1998xd,Israel:2004ir}. Its transverse part
is T-dual to $[SU(2)_k/U(1) \times SL(2,\mathbb{R})_k/U(1)]/\mathbb{Z}_k$.

D-branes in this background have been constructed in~\cite{Eguchi:2004ik,Israel:2005fn}.
First the suspended D-branes are made, in the T-dual background, of a D0-brane
of the cigar $SL(2,\mathbb{R})/U(1)$ sitting at the tip and a D2-brane of $SU(2)/U(1)$ with
the shape of a disk. While the former carries no label, the latter has one parameter
$\hat \jmath =0,1/2,\ldots k/2-1$ giving its
length in the fivebranes geometry. For the flat $\mathbb{R}^{5,1}$ part of the background we choose NNNNDD boundary
conditions in type IIB, giving a non-\textsc{bps} D4-brane stretched between two fivebranes, i.e. an S-type brane
(see the right picture of fig~\ref{ringpic}).
The annulus amplitude for open strings  between two identical S-type branes
reads:
\begin{multline}
\mathcal{A}^{\rm s}_{\hat \jmath \hat \jmath} =  \sum_{a, \upsilon_i \in \zi_2}(-)^a
\int \frac{\di t}{2t} \frac{\Theta_{a+2\upsilon_1,2}\, \Theta_{a+2\upsilon_2,2} }{(8\pi^2\alpha' t)^2\eta^6}\ \times \\
\times
\sum_{\ell =0}^{\mathrm{min} (2{\hat \jmath},k-2{\hat \jmath})}
 \sum_{m \in \zi_{2k}} C^{\ell\, (a+2\upsilon_3)}_{m} (it)\,
C\!h_\mathbb{I}^{(a+2\upsilon_4)} \left(\frac{m-a}{2};it\right)\, .
\label{sk}
\end{multline}
The supersymmetric $SU(2)/U(1)$ characters $C^{\ell\, (s)}_{m}$ and $SL(2,\mathbb{R})/U(1)$ extended identity
character $C\!h_\mathbb{I}^{(s)} (r)$ can be found in app.~\ref{charapp}.
One has a {\it universal} open string
tachyon for $\ell=m=\upsilon_i=0$ in the \textsc{ns} sector ($a=0$), of mass squared $M^2 = -\nicefrac{1}{2\alpha'}$.

A second type of brane of interest is made in the T-dual of a D0-brane of $SL(2,\mathbb{R})/U(1)$ and a D1-brane
of SU(2)/U(1), with similar boundary conditions for $\mathbb{R}^{5,1}$. 
In the $\zi_k$ orbifold theory these boundary conditions are not compatible hence the
generalized \textsc{gso} projection does not act in the open string sector.
It corresponds in the fivebranes geometry  to a disk at the center of the ring of
radius $ R \sin  \frac{\pi}{k}(2\hat \jmath+1)$~\cite{Israel:2005fn},
i.e. a G-type brane (of nonzero radial extension), see
left picture of fig.~\ref{ringpic}. The annulus amplitude for identical branes is:
\begin{multline}
\mathcal{A}^{\rm g}_{\hat \jmath \hat \jmath} = \frac{1}{2}\!\sum_{a,b, \upsilon_i \in \zi_2}
(-)^{a+b(1+\sum_i \upsilon_i)}
\int \frac{\di t}{2t} \frac{\Theta_{a+2\upsilon_1,2}\Theta_{a+2\upsilon_2,2} }{(8\pi^2 \alpha' t)^2\eta^6}
\ \times \\
\times
\sum_{\ell =0}^{\mathrm{min} (2{\hat \jmath},k-2{\hat \jmath})}
 \sum_{m \in \zi_{2k}}  C^{\ell\, (a+2\upsilon_3)}_{m} (it)
\sum_{r \in \zi_k} C\!h_\mathbb{I}^{(a+2\upsilon_4)} (r;it) \, .
\label{gk}
\end{multline}
The $\nicefrac{1}{2}$ factor comes with the fermionic \textsc{gso} projection as this brane
has the correct dimensionality (odd in type IIB). However states with $r \neq (m-a)/2$ break space-time supersymmetry.
The spectrum includes a geometric tachyon (not projected out by the fermionic \textsc{gso})
for $\ell=r=0$, $\upsilon_3=1$ and $m=2$ of mass squared
$M^2=-\nicefrac{1}{k\alpha'}$~\cite{Eguchi:2004ik}.

\FIGURE{
\centering
\epsfig{file=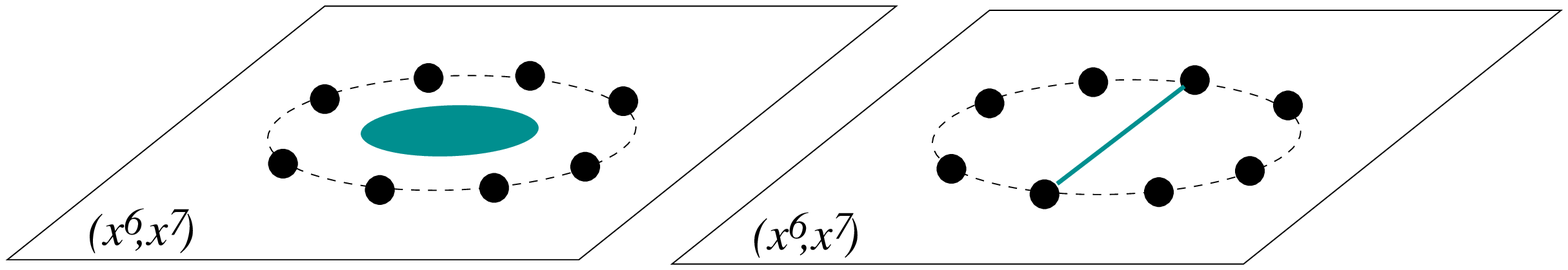,width=120mm}
\caption{\it G-type  (left) and S-type (right) D-branes inside a ring of NS5-branes}
\label{ringpic}
}
\section{D-branes near a pair of fivebranes}
\label{pair}
We now specialize the discussion to $k=2$, i.e. a pair of fivebranes.
At first glance we see that the tachyon masses coincide: $M^2=-\nicefrac{1}{2\alpha'}$.

Let us look in more detail at annulus amplitude for both types of branes.
At level $k=2$ the super-coset $SU(2)_k/U(1)$  simplifies dramatically as it
contains only the identity field. From the defining relation of  coset characters
\begin{equation*}
\chi^j \Theta_{s,2} =  \sum_{m \in \zi_{2k}} C^{j\, (s)}_{m} \Theta_{m,k}\, ,
\end{equation*}
one sees that for $k=2$ one gets the constraint $m=s$ (in addition to $j=0$).
For an S-type brane, the annulus amplitude, eqn.~(\ref{sk}), becomes:
\begin{equation}
\mathcal{A}^{\rm s}_{k=2} =  \sum_{a, \upsilon_i \in \zi_2}(-)^a
\int \frac{\di t}{2t} \frac{\Theta_{a+2\upsilon_1,2}\Theta_{a+2\upsilon_2,2} }{(8\pi^2 \alpha' t)^2\eta^6}
\, C\!h_\mathbb{I}^{(a+2\upsilon_4)} (\upsilon_3)\, ,
\label{s2}
\end{equation}
where the tachyon occurs for $\upsilon_i=0, \, \forall i$ in the \textsc{ns} sector.

For the G-type brane annulus amplitude the  $m=s$ constraint has no effect on the remaining
degrees of freedom, therefore  one can simply replace in eqn.~(\ref{gk}) the sum of $SU(2)/U(1)$
characters  $\sum_m C^{\ell (s)}_m$ by one:
\begin{equation}
\mathcal{A}^{\rm g}_{k=2} =  \frac{1}{2} \sum_{a,b, \upsilon_i \in \zi_2}
(-)^{a+b(1+\sum\limits_{i=1}^{4} \upsilon_i)}
\int \frac{\di t}{2t} \frac{\Theta_{a+2\upsilon_1,2}\Theta_{a+2\upsilon_2,2} }{(8\pi^2 \alpha' t)^2\eta^6}
\sum_{r \in \zi_2} C\!h_\mathbb{I}^{(a+2\upsilon_4)} (r) \, .
\label{g2}
\end{equation}
However, the  fermionic $\zi_2$ label  $\upsilon_3$ of $SU(2)/U(1)$ still appears
in the phase of the fermionic \textsc{gso} projection. It gives
an overall factor in the annulus amplitude
\begin{equation}
\frac{1}{2} \sum_{\upsilon_3=0}^1 (-)^{\upsilon_3\, b} = \delta_{b,0 \mod 2}\, .
\end{equation}

Therefore only the $b=0$ sector (i.e. the untwisted \textsc{ns} and \textsc{r} sectors) survives for $k=2$.
It shows that the annulus amplitudes~(\ref{s2},\ref{g2}) are indeed identical once we identify $\upsilon_3$
in eqn.~(\ref{s2}) with $r$ in eqn.~(\ref{g2}).

\section{Discussion}
In this note we compared  compact
G-type branes (would-be \textsc{bps}-branes with a geometric tachyon due to the background)
and S-type branes (stretched non-\textsc{bps} branes) inside a ring of fivebranes, 
using boundary worldsheet \textsc{cft}.
In the limit where the ring consists in two fivebranes, we have shown that the S-type and G-type branes give
the same annulus amplitude.

Compared to the setup of~\cite{Kutasov:2004ct,Sen:2007cz} the space transverse to
the fivebranes is not compactified on a circle. For this particular sector of D-branes that are either
stretched or halfway between a pair of fivebranes it is irrelevant. However the transverse circle
should be taken into account if one includes also U-type branes wrapping the circle in the discussion.

Firstly, the matching of one-loop open string amplitudes
implies that both branes have the same open string tachyon, supporting
the conjecture made in~\cite{Sen:2007cz}. The decay of this tachyon has been studied in~\cite{Israel:2006ip} using
\textsc{cft} methods. It was found that, as in flat space, the outcome of the process is a dust of very massive
closed strings.

Secondly, because the whole annulus amplitudes are identical, by channel duality one finds that
the brane couplings to closed strings are the same up to a phase. It can be checked  at
the level of the boundary states that the couplings are actually identical, see app.~\ref{appbdy}.
In particular having identical couplings to the graviton means that the G-type and the S-type branes have the
same tension.

Similar correspondences between universal and geometric open string tachyons near a pair of NS5-branes
exist for  non-compact branes. For instance, the non-\textsc{bps} "D-rays"
of~\cite{Israel:2005fn},  identified in the large $k$ limit as a pair of semi-infinite D1-branes ending on
the fivebranes from outside the ring, and the second class of G-type D4-branes 
(sect.~5.2 in~\cite{Israel:2005fn}) with $\hat \jmath =0$, 
filling the $x^{8,9}$ plane at $x^6=x^7=0$  (the NS5-branes are spread in the $x^{6,7}$ plane),
become equivalent for $k=2$. It would be interesting to investigate these aspects further.

These results illustrate how space-time geometry is modified at large curvatures in string theory,
whenever all  $\alpha'$ corrections are taken into account.

\acknowledgments
I thank Ashoke Sen for very useful comments about the first version of this work.

\appendix

\section{$\mathcal{N}=2$ characters}
\label{charapp}
\subsection*{Free fermions}
In order to write  free-fermion characters we use $\vartheta \oao{a}{b} (\tau,\nu )
= \sum_{n \in \zi} q^{\frac{1}{2} (n+\frac{a}{2})^2}
e^{2 i \pi (n+\frac{a}{2})(\nu+\frac{b}{2})}$.
It is convenient to split the R and NS sectors
according to the fermion number mod 2:
\begin{equation}
\begin{array}{cccc}
\frac{1}{2\eta} \left\{ \vartheta \oao{0}{0} - \vartheta \oao{0}{1} \right\} & =
\frac{\Theta_{0,2}}{\eta} &
 \frac{1}{2\eta} \left\{ \vartheta \oao{0}{0} + \vartheta \oao{0}{1} \right\} & =
\frac{\Theta_{2,2}}{\eta}  \\
\frac{1}{2\eta} \left\{ \vartheta \oao{1}{0} - i\vartheta \oao{1}{1} \right\} & =
\frac{\Theta_{1,2}}{\eta} &
\frac{1}{2\eta} \left\{ \vartheta \oao{1}{0} + i\vartheta \oao{1}{1} \right\} & =
\frac{\Theta_{3,2}}{\eta}
\end{array}
\end{equation}
in terms of the theta functions of $\hat{\mathfrak{su}} (2)$
at level $2$:
$\Theta_{m,k} (\tau,\nu) = \sum_{n \in \zi}
q^{k\left(n+\frac{m}{2k}\right)^2}
e^{2i\pi \nu k \left(n+\frac{m}{2k}\right)}.$
The modular transformation property of these $U(1)_2$ characters is then:
\begin{equation}
\frac{\Theta_{s,2} (-1/\tau , \nu/\tau )}{\eta (-1/\tau  )}  = \frac{1}{2}
\, e^{i\pi \nu^2 /\tau}
\sum_{s' \in \zi_4}
e^{-\frac{i\pi ss'}{2}} \frac{\Theta_{s',2} (\tau , \nu)}{\eta (\tau )}  .
\end{equation}
\boldmath
\subsection*{$\cn =2$ minimal models}
\unboldmath
The characters of the $\cn =2$ minimal model, i.e.  the supersymmetric gauged \textsc{wzw} model
$SU(2)_k / U(1)$,  are determined implicitly through the
identity:
\begin{equation}
\sum_{m \in \zi_{2k}} \mathcal{C}^{j\ (s)}_{m}  \Theta_{m,k} = \chi^{j}
\Theta_{s,2}\, ,
\end{equation}
where $\chi^j$ is a character of $SU(2)$ at level $k-2$.
They are labeled by  $(j,m,s)$, corresponding  to
primaries of the coset $[SU(2)_{k-2}\times U(1)_2]/U(1)_{k}$.
The following identifications apply:
\begin{equation}
(j,m,s) \sim (j,m+2k,s)\sim
 (j,m,s+4)\sim
 (k/2-j-1,m+k,s+2)
\end{equation}
as  the selection rule $2j+m+s =  0  \mod 2$.
The weights of the primaries states are:
\begin{equation}
\begin{array}{cccccc}
h &=& \frac{j(j+1)}{k} - \frac{n^2}{4k} + \frac{s^2}{8} \ & \text{for} & \ -2j \leqslant n-s \leqslant 2j \\
h &=& \frac{j(j+1)}{k} - \frac{n^2}{4k} + \frac{s^2}{8} + \frac{n-s-2j}{2}
\ & \text{for} & \ 2j \leqslant n-s \leqslant 2k-2j-4 \\
\end{array}
\end{equation}
We have the following modular S-matrix for these characters:
\begin{equation}
S^{j\, m\, s}_{\quad j' \, m' \, s'} = \frac{1}{2k} \sin \pi
\frac{(1+2j)(1+2j')}{k} \ e^{i\pi \frac{mm'}{k}}\ e^{-i\pi ss'/2}.
\end{equation}
\subsection*{Supersymmetric $SL(2,\mathbb{R})/U(1)$}
The characters of the $SL(2,\mathbb{R})/U(1)$ super-coset
at level $k$ come in different categories corresponding to
irreducible unitary representations of  $SL(2,\mathbb{R})$.
The \emph{continuous representations} correspond to $j = 1/2 + iP$,
$P \in \mathbb{R}^+$. Their characters are denoted by
 $ch_c (p,m) \oao{a}{b}$,
where the $U(1)_R$ charge of the primary is $Q=2m/k$.
The \emph{discrete representations}, of characters $ch_d (j,r) \oao{a}{b}$,
have $1/2 < j < (k+1)/2$, with $U(1)_R$  charge  $Q=2(j+r+a/2)/k$,
$r\in \zi$. The \emph{identity representation} primaries have  $U(1)_R$ charge
$Q=(2r+a)/k$, $r \in \mathbb{Z}$,  and the  characters read~:
\begin{equation}
ch_\mathbb{I} (r;\tau,\nu) \oao{a}{b} =  \frac{(1-q)\
  q^{\frac{-1/4+(r+a/2)^2}{k}}
e^{2i\pi\nu \frac{2r+a}{k}}}{\left( 1+(-)^b \,
e^{2i\pi \nu} q^{1/2+r+a/2} \right)\left( 1+(-)^b \, e^{-2i\pi \nu}
q^{1/2-r-a/2}\right)} \frac{\vartheta \oao{a}{b} (\tau, \nu)}{\eta^3 (\tau)}.
\label{idchar}
\end{equation}
One can define characters labeled by a quantum number $s  \in \zi_4$,
following the method we used for  free fermions. The spectrum of NS primaries
for the identity representation is as follows.
The identity operator is $| 0\rangle_\textsc{ns} \otimes |r=0\rangle_\textsc{sl(2,r)}$,
in the sector $s=0$. The other primaries, in the sector $s=2$, have weights $
h= \frac{r^2}{k} +|r| - \frac{1}{2}
$.

\emph{Extended characters} are defined for $k$ integer by summing
over $k$ units of spectral flow.  The extended identity characters are:
\begin{equation}
  Ch_\mathbb{I}^{(s)} (r;\tau)= \sum_{w \in \zi} ch_\mathbb{I}^{(s)} \left( r+kw;\tau \right)
   \quad \text{with} \qquad r \in \zi_k
\end{equation}
Their modular transformation involve only a discrete set
of $\mathcal{N}=2$ charges:
\begin{multline}
Ch_{\mathbb{I}}^{(a+2\upsilon)} (r ;-1/\tau )
= \frac{1}{k} \sum_{a', \upsilon' \in \zi_2} e^{-\frac{i \pi}{2} (a+2\upsilon) (a'+2\upsilon')}\
\times \\ \times \
\left[
\int_{0}^{\infty} \di P'  \sum_{m' \in \zi_{2k}} e^{-\frac{2i\pi}{k} (r+\frac{a}{2}) m'}
\frac{ \sinh 2\pi P'  \sinh \frac{2\pi P'}{k}}{\cosh 2\pi P' + \cos \pi (m'-a')}
 Ch_{c}^{(a'+2\upsilon')} (P',\frac{m'}{2};\tau)  \right. \\
\left.  +  \sum_{2j'-1=1}^{k-1}\  \sum_{r' \in \zi_k}
\sin \frac{\pi (1-2j')}{k} \,  e^{-\frac{4i \pi}{k} (j'+r'+\frac{a'}{2})(r+\frac{a}{2})}
\, Ch_{d}^{(a'+2\upsilon')} (j',r' ; \tau) \right]
\end{multline}

\section{Boundary states}
\label{appbdy}
\subsection*{S-type brane boundary state}
We consider the boundary state for a type IIB S-type D4-brane in
$\mathbb{R}^{5,1} \times [SU(2)_k/U(1) \times SL(2,\mathbb{R})_k/U(1)]/\mathbb{Z}_k$.
The coefficients of the
Ishibashi states are given by  the  one-point functions~(4.7)
of~\cite{Israel:2005fn}\footnote{Compared to~\cite{Israel:2005fn}  the brane has a different dimensionality.
The fivebranes are stretched along $x^{0,\ldots,5}$  and distributed in the $x^{6,7}$ plane,
the D4-brane is stretched along $x^{0,1,2,3}$ and one direction in the $x^{6,7}$ plane.
} (see also~\cite{Eguchi:2003ik}),
keeping only couplings to the NS-NS sector with an extra $\sqrt{2}$ factor
to satisfy the Cardy condition (as non-\textsc{bps} branes in flat space). Its labels $(\hat \jmath, \hat m)$
(with $0\leqslant 2\hat \jmath \leqslant k-2$ and
$\hat m \in \mathbb{Z}_{2k}$) give the position of the two fivebranes on which the D-brane ends; we choose
$\hat \jmath =0$  (i.e. shortest length). As in flat space the brane has fermionic labels
$\{\hat s_i \in \mathbb{Z}_4 , i=1,\ldots,4\}$, and its position is  $(\hat x^4,\hat x^5)$.
The couplings to $SL(2,\mathbb{R})/U(1)$  continuous representations are
(the discrete ones follow from analyticity):
\begin{multline}
|{\rm S} \, (\hat x^\mu, \hat \jmath=0, \, \hat m ,\, \{\hat s_i \})\, \rangle_k = \mathcal{T}
 \frac{\sqrt{2}}{k}
\int\frac{d^2 p}{(2\pi)^2} \ e^{i (p_4 \hat x^4 + p_5 \hat x^5)} \sum_{\upsilon_i \in \zi_2} \frac12
\sum_{b=0}^1 (-)^{b(1+\sum_i \upsilon_i)} \ \times \\ \times \
e ^{i\pi \sum_{i=1}^4 \upsilon_i
\hat{s}_i}
|p_i , s_1=-\bar s_1 =2\upsilon_1,s_2= -\bar s_2= 2\upsilon_2\rrangle_{\rm flat}  \\  \otimes \
\sum_{2j'=0}^{k-2}\sum_{m \in \mathbb{Z}_{2k}} \delta_{2j'+m,0 \mod 2}\
e^{-\frac{i\pi}{k} m \hat m}
\sqrt{\sin \frac{\pi}{k} (2j'+1)} 
| j',m,-m,s_3=\bar s_3=2\upsilon_3\rrangle_{\mathfrak{su(2)/u(1)}} \\ \otimes \
\int_0^\infty \!\! dP\ \nu_k^{-iP} \sum_{w \in \mathbb{Z}} \frac{\Gamma(\hl+iP +\frac{m}{2}-\upsilon_4+kw)\,
\Gamma(\hl+iP -\frac{m}{2}+\upsilon_4-kw)}{\Gamma (2iP)\, \Gamma (1+2iP/k)}\  \times \\ \times\
|j=1/2+iP, m/2+kw,-m/2-kw,s_4=-\bar s_4 = 2\upsilon_4\rrangle_{\mathfrak{sl(2)/u(1)}}
\end{multline}
with the normalization $\mathcal{T}= (\nicefrac{2}{\alpha'})^{\nicefrac{1}{2}} /8\pi$ from $\mathbb{R}^{5,1}$.
The sum over $b=0,1$ implements the closed string fermionic \textsc{gso} projection. We denote by $|\star \rrangle$ the Ishibashi
states for flat space with ${\rm NNDD}$ boundary conditions, supersymmetric
$SU(2)/U(1)$ and $SL(2,\mathbb{R})/U(1)$ with B-type
boundary conditions respectively.

Let us now choose $k=2$. As we saw in the bulk of the paper, for the $SU(2)/U(1)$ coset we have only
the identity, i.e.   $j'=0$ and $m=s_3=2\upsilon_3$. It gives the boundary state:
\begin{multline}
|{\rm S} \, (\hat x^\mu,  \hat m ,\, \{\hat s_i \})\, \rangle_2 = \mathcal{T} \frac{1}{\sqrt{2}}
\int \frac{d^2 p}{(2\pi)^2}\ e^{i (p_4 \hat x^4 + p_5 \hat x^5)}
 \sum_{\upsilon_i \in \zi_2 } \frac12 \sum_{b=0}^1 (-)^{b(1+\sum_i \upsilon_i)}
e ^{i\pi \sum_{i=1}^4 \upsilon_i
\hat{s}_i}\times \  \\ \times \ |p_i , s_1=-\bar s_1=2\upsilon_1,s_2 =-\bar s_2 = 2\upsilon_2\rrangle_{\rm flat}  \  \otimes \
e^{-i\pi \upsilon_3 \hat m}\
| 0\rrangle_{\mathfrak{su(2)/u(1)}} \\ \otimes \
\int_0^\infty \!\! dP\ \nu_2^{-iP} \sum_{w \in \mathbb{Z}} \frac{\Gamma(\hl+iP +\upsilon_3-\upsilon_4+2w)\,
\Gamma(\hl+iP -\upsilon_3+\upsilon_4-2w)}{\Gamma (2iP)\, \Gamma (1+iP)}\  \times \\ \times\
|j=1/2+iP, \upsilon_3+2w,-\upsilon_3-2w,s_4=-\bar s_4=2\upsilon_4\rrangle_{\mathfrak{sl(2)/u(1)}}
\label{stypek2}
\end{multline}
\subsection*{G-type brane boundary state}
We consider the G-type D5-brane, still in type IIB. The brane has the shape of a disc in the $x^{6,7}$ plane, whose radius is parameterized
by $\hat \jmath$. We choose $\hat \jmath =0 $, i.e. a
"fuzzy" D3-brane at the center of the ring.  The coefficients  of the Ishibashi states are obtained from the  one-point
function~(5.2) of~\cite{Israel:2005fn}. One gets the G-type D3-brane boundary state as~\cite{Eguchi:2004ik,Israel:2005fn}:\footnote{In~\cite{Israel:2005fn} the
Ishibashi states with $\epsilon=1$ were missing. They appear because
the orbifold is of order $k$ and not $2k$ (as  the B-brane of SU(2) discussed in~\cite{mms}). We
thank A.~Sen for pointing out this to us.}
\begin{multline}
|{\rm G} \, (\hat x^\mu,\hat \jmath=0,\eta,  \{\hat s_i \})\, \rangle_k =  \mathcal{T} \frac{1}{\sqrt{k}}
\int \frac{d^2 p}{(2\pi)^2}\ e^{i (p_4 \hat x^4 + p_5 \hat x^5)}
\sum_{\upsilon_i \in \zi_2} \frac12 \sum_{a,b=0}^1 (-)^{a+b(1+\sum_i \upsilon_i)}
  \ \times \\ \times \ e ^{\frac{i\pi}{2} \sum_{i=1}^4
(a+2\upsilon_i)
\hat{s}_i}\
|p_i , s_1=-\bar s_1=a+2\upsilon_1,s_2 =-\bar s_2= a+2\upsilon_2\rrangle_{\rm flat}  \\  \otimes \
\sum_{2j'+1=1}^{k-1} \sum_{\epsilon \in \zi_2} \eta^\epsilon
\sqrt{\sin \frac{\pi}{k} (2j'+1)}\ \delta_{2j'+k\epsilon+a,0 \mod 2}
| j',0,  0,
s_3=-\bar s_3=a+2(\upsilon_3+\epsilon)\rrangle_{\mathfrak{su(2)/u(1)}} \\ \otimes \
\int_0^\infty \!\! dP\ \nu_k^{-iP} \sum_{w \in \mathbb{Z}} \frac{\Gamma(\hl+iP+\frac{k\epsilon}{2} -\upsilon_4+kw)\,
\Gamma(\hl+iP -\frac{k\epsilon}{2}+\upsilon_4-kw)}{\Gamma (2iP)\, \Gamma (1+2iP/k)}\ \times \\ \times\
|j=1/2+iP,k(w+\epsilon/2),-k(w+\epsilon/2),
s_4=-\bar s_4 = a+2\upsilon_4\rrangle_{\mathfrak{sl(2)/u(1)}}
\end{multline}
where now the $SU(2)/U(1)$ Ishibashi states correspond to A-type boundary conditions. The
label $\eta = \pm 1$ is a $\mathbb{Z}_2$-valued Wilson line~\cite{mms}. 

Let us consider the case $k=2$. The states that survive have $j'=0$ and $a+2(\upsilon_3+\epsilon)=0$,
hence only the NS-NS sector (i.e. $a=0$) together with the constraint $\epsilon=-\upsilon_3 \mod 2$. 
One gets the same  boundary state as for the S-type brane, eqn.~(\ref{stypek2}), once
we identify the brane labels as $\eta=e^{i\pi\hat m }$.

\end{document}